\documentclass[12pt,numbers,sort&compress]{elsarticle}
\usepackage{amsmath}
\usepackage{amssymb}
\usepackage{booktabs}
\usepackage{hyperref}
\usepackage{xcolor}
\usepackage{listings}
\usepackage{enumitem}
\usepackage{tabularx}
\usepackage{longtable}
\usepackage{array}

\usepackage[a4paper,margin=2.2cm]{geometry}

\newcommand{\add}[1]{#1}

\makeatletter
\def\ps@pprintTitle{%
  \let\@oddhead\@empty
  \let\@evenhead\@empty
  \let\@oddfoot\@empty
  \let\@evenfoot\@empty}
\makeatother

\newcounter{bla}

\hypersetup{
  colorlinks=true,
  linkcolor=blue,
  citecolor=blue,
  urlcolor=blue
}

\lstdefinestyle{ini}{
  basicstyle=\ttfamily\small,
  breaklines=true,
  columns=fullflexible,
  frame=single,
  backgroundcolor=\color{gray!5}
}

\lstdefinestyle{json}{
  basicstyle=\ttfamily\small,
  breaklines=true,
  columns=fullflexible,
  frame=single,
  backgroundcolor=\color{gray!5}
}

\lstdefinestyle{prompt}{
  basicstyle=\ttfamily\footnotesize,
  breaklines=true,
  columns=fullflexible,
  frame=single,
  backgroundcolor=\color{gray!5},
  xleftmargin=0.02\textwidth,
  xrightmargin=0.02\textwidth,
  aboveskip=0.6em,
  belowskip=0.6em
}

\begin{document}

\begin{frontmatter}

\title{\texttt{EasyScan\_HEP~2}: LLM-Agent Parameter-Scan Workflows in High Energy Physics}

\author[inst1]{Yang Xiao}
\ead{xiaoyangphy@gmail.com}
\author[inst1]{Yuanfang Yue}
\ead{yueyuanfang@htu.edu.cn}
\author[inst1]{Yang Zhang\corref{cor1}}
\ead{zhangyang2025@htu.edu.cn}
\cortext[cor1]{Corresponding author}
\affiliation[inst1]{organization={School of Physics, Henan Normal University}, city={Xinxiang 453007}, country={China}}

\begin{abstract}

\add{Large-language-model~(LLM)} agents are beginning to reshape the preparation and steering of computational workflows in high-energy physics phenomenology. To accommodate this change, we upgrade \texttt{EasyScan\_HEP} to make the construction of parameter-scan configuration files more accessible to LLM-agent assistance.
\texttt{EasyScan\_HEP~2} \add{exposes command-line and machine-readable interfaces for LLM-agent workflows}, allowing an assistant to translate natural language requests into an explicit \texttt{.ini} configuration that defines the scan method, external-program workflow, constraints, and outputs. 
The resulting configuration can be inspected through a local Web \add{interface}.
The framework also supports \add{LLM-agent-guided} extension to new scan methods, as illustrated by the integration of \texttt{BESTFIT}, \texttt{EMCEE}, and \texttt{DYNESTY}.
In this way, \texttt{EasyScan\_HEP~2} adapts \add{parameter-scan workflows to LLM-agent use} while preserving reproducibility, transparency, and user control.
\vspace{1em}

\end{abstract}


\end{frontmatter}

\newpage

\section{Introduction}

Artificial intelligence \add{(AI)} and machine learning have become an important part of modern high-energy physics (HEP).
They are widely used in event reconstruction, object identification, jet tagging, anomaly detection, fast simulation, statistical inference, and phenomenological studies~\cite{Albertsson:2018maf, Guest:2018yhq,Feickert:2021ajf,Karagiorgi:2021ngt,Plehn:2022ftl,Richmond:2025lzg,Qu:2019gqs,Paganini:2017dwg,Brehmer:2019xox,Brehmer:2018kdj,Searle:2025cnj,Bian:2025yfj}.
More recently, \add{agents built on large language models (LLMs)} have begun to extend AI assistance from isolated inference tasks to the orchestration of scientific workflows, including code generation, tool invocation, structured context management, and human-in-the-loop analysis~\cite{Menzo:2025cim,Gendreau-Distler:2025fsj,Moreno:2026mqk,Faroughy:2026dkj,Qiu:2026iby,Agrawal:2026lvg,Lucente:2026kgh,Wang:2026jjn,Costa:2026oew,Alexander:2026lpw}.

Parameter-space exploration remains one of the central computational tasks in HEP phenomenology.
Studies of physics beyond the Standard Model \add{(BSM)} often require scanning multidimensional model parameters under collider, flavor, precision, dark matter~(DM), and cosmological constraints.
A number of public tools have been developed for this purpose, including early supersymmetry fitting frameworks such as \texttt{Fittino}~\cite{Bechtle:2004pc} and \texttt{SFitter}~\cite{Lafaye:2004cn}, global-inference frameworks such as \texttt{GAMBIT}~\cite{GAMBIT:2017yxo} and \texttt{ScannerBit}~\cite{Martinez:2017lzg}, constraint-combination tools such as \texttt{HEPfit}~\cite{DeBlas:2019ehy}, model-oriented scanners such as \texttt{ScannerS}~\cite{Muhlleitner:2020wwk}, BSM scanning tools such as \texttt{xBIT}~\cite{Staub:2019xhl}, \texttt{BSMArt}~\cite{deSouza:2026jww,Goodsell:2023iac}, and \texttt{hep-aid}~\cite{Diaz:2024sxg}, and workflow-oriented frameworks such as \texttt{EasyScan\_HEP}~\cite{Shang:2023gfy} and \texttt{Jarvis-HEP}~\cite{Guo:2026kfy}.
In parallel, AI-assisted strategies have been introduced to improve the efficiency of such scans, for example through machine-learning-guided sampling, active learning, surrogate modelling and deep-learning-assisted exploration
\cite{Ren:2017ymm,Caron:2019xkx,deSouza:2022uhk,Hammad:2022wpq,DarkMachinesHighDimensionalSamplingGroup:2021wkt,AbdusSalam:2020rdj,Hammad:2024tzz,Zeng:2026zqn}.  

\add{As LLM-agent workflows become practical for scientific workflow orchestration,} parameter scans need explicit and machine-readable workflow descriptions beyond the sampling strategy itself.
For many HEP scans, a substantial part of the work lies not only in choosing a sampler, but also in preparing the computational workflow: connecting external physics programs, modifying input cards point by point, reading output observables, defining likelihoods and constraints, and storing the scan setup together with its outputs. 
This is the motivation for \texttt{EasyScan\_HEP~2}.
Rather than improving the scan engine, the upgraded framework lets AI assist the configuration layer.
The scientific content of the scan remains encoded in an explicit \texttt{EasyScan\_HEP} configuration file, which specifies the scan method, input parameters, external programs, input-output mappings, constraints, plots, and result folder.
The file can be generated or revised with AI assistance and then executed by the same \texttt{EasyScan\_HEP} backend, while the command-line tools and the local Web \add{user interface (UI)} expose the configuration before execution.

The paper is organized as follows.
Section~2 gives a quick start for using \texttt{EasyScan\_HEP~2} through an agent skill.
Section~3 summarizes the \add{configuration-layer architecture} and the main software upgrades, and presents \add{a controlled evaluation of \texttt{EasyScan\_HEP} workflows}.
Section~4 describes command-line and Web UI use without an autonomous agent.
Section~5 discusses \add{LLM-agent-guided} extension to future samplers, with \texttt{BESTFIT}, \texttt{EMCEE}, and \texttt{DYNESTY} as examples.
Section~6 presents a phenomenological scan of the \add{scalar singlet model} involving relic-density and phase-transition calculations.
Section~7 concludes.

\section{Quick start}
\label{sec:agent-skill}

The entry point for \add{LLM-agent use} is a dedicated \texttt{EasyScan\_HEP~2} agent skill.
This skill is separate from the \texttt{EasyScan\_HEP} package: it contains
\add{instructions for LLM-agent preparation and checking of configuration files}, while the
package installed in Section~\ref{sec:agent-oriented-upgrades} contains the
scanner itself.  The skill is distributed from
\begin{verbatim}
https://github.com/Contract-Mediated-Agent/easyscan-skill.git
\end{verbatim}
In an \add{LLM-agent environment} that supports installable skills, such as Codex-style
workflows, the user may install the skill by
\begin{verbatim}
mkdir -p ~/.codex/skills
git clone https://github.com/Contract-Mediated-Agent/easyscan-skill.git \
    ~/.codex/skills/easyscan-hep
\end{verbatim}
\add{The skill is versioned separately from the \texttt{EasyScan\_HEP} package
and targets the \texttt{EasyScan\_HEP~2.0} interfaces described in this work.}
After restarting the agent if necessary, this skill can be invoked explicitly or selected by the agent when the user asks for an \texttt{EasyScan\_HEP} scan.

As a quick-start example, one can reproduce the example Listing~1 in the \texttt{EasyScan\_HEP~1} documentation~\cite{Shang:2023gfy}, by asking the agent as follows.
For this particular prompt, the user should first download or clone a local \texttt{EasyScan\_HEP} source tree, because the example refers to the bundled \texttt{utils/TestFunction.py} and its companion input/output files.
\add{If the prompt is used outside such a source tree, the skill first checks whether these bundled files are available and otherwise asks the user to obtain the example resources before generating a runnable configuration.}
\begin{quote}
\sffamily{
[\$easyscan-hep] Scan \texttt{x} from 0 to 3.14 and \texttt{y} from -3.14 to 3.14,
randomly with 100 total points. Put results in
\texttt{example\_random}. Run \texttt{./TestFunction.py} in \texttt{utils/}. Write \texttt{x} and \texttt{y} to
\texttt{utils/TestFunction\_input.dat} at row 1 columns 1 and 2. Read \texttt{f} from
\texttt{utils/TestFunction\_output.dat} at row 1 column 2. Add \texttt{Gaussian} \texttt{f} = 1 +/- 0.2.
Make \texttt{Color} plots for \texttt{x}, \texttt{y}, \texttt{f} and \texttt{x}, \texttt{y}, \texttt{Chi2}.
}
\end{quote}
If a working \texttt{EasyScan\_HEP} installation is not available, the skill guides the user through the installation procedure described in the next section.
Then, the user's request is translated into a complete \texttt{.ini} file.
When the request is not sufficiently specified, the skill turns the missing items into explicit follow-up questions for the user.
\add{This conservative behavior is deliberate.  In LLM-agent workflows, a
substantial part of the cost comes from human inspection of the generated setup,
so the skill is designed to reduce unchecked guesses rather than to maximize
autonomous completion.  For example, if the user does not provide the location
of an external program, the skill asks for the path instead of searching the
local file system and choosing a candidate.  If several local versions of the
same program are present, such a guess may select the wrong executable.  Since
this information is scientifically and operationally important and must be
checked by the user in any case, asking for it explicitly reduces the risk of an
unnoticed wrong choice.}

The skill provides a \add{natural-language entry point} to \texttt{EasyScan\_HEP}.
A user can describe the intended scan in natural language and let the \add{LLM-agent} prepare, check, and revise the configuration, without learning the full command-line syntax or the detailed configuration rules.

\section{\texorpdfstring{\add{Upgrades for LLM-agent workflows}}{Upgrades for LLM-agent workflows}}
\label{sec:agent-oriented-upgrades}

\begin{figure}[t]
\centering
\includegraphics[width=0.98\textwidth]{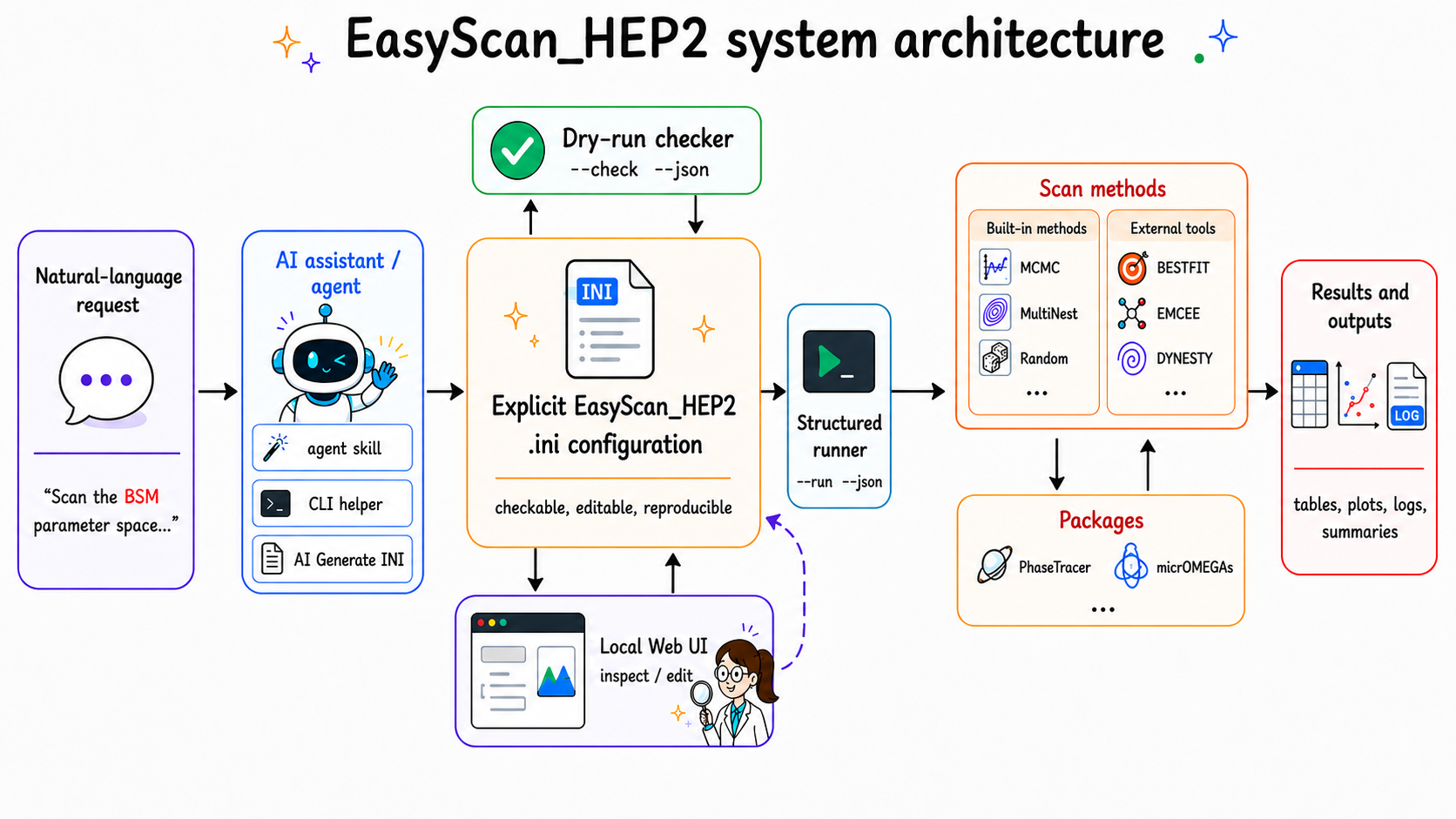}
\caption{\add{Architecture of \texttt{EasyScan\_HEP~2} for LLM-agent workflows}.}
\label{fig:architecture}
\end{figure}

We now describe the software upgrades that make this workflow possible.
The basic scientific contract of \texttt{EasyScan\_HEP} is unchanged, and the upgrade instead adds \add{machine-readable interfaces} around this original configuration workflow.
Figure~\ref{fig:architecture} summarizes how these interfaces are organized around the same checked \texttt{.ini} file.

\subsection{Installable command-line entry point}

A first practical change is that \texttt{EasyScan\_HEP~2} is organized as an installable Python package. It can be installed from the GitHub repository,
\begin{verbatim}
python3 -m pip install \
    git+https://github.com/phyzhangyang/EasyScan_HEP.git
\end{verbatim}
or directly from a local source tree,
\begin{verbatim}
python3 -m pip install .
\end{verbatim}
After installation, the command
\begin{verbatim}
easyscan config.ini
\end{verbatim}
is available from any working directory.
The original source-tree usage remains available for development.

The installed command resolves relative paths in a configuration file from the directory where \texttt{easyscan} is launched.
For this reason, an \add{LLM-agent} should generate the \texttt{.ini} file in the intended launch directory, or use explicit paths when the user requests them.

\subsection{Dry-run configuration checker}

Another \add{machine-readable} upgrade is the configuration checker.
A user or \add{LLM-agent} can check a configuration file without launching any scan point:
\begin{verbatim}
easyscan --check config.ini
\end{verbatim}
For \add{LLM-agent use}, the same check can return machine-readable output:
\begin{verbatim}
easyscan --check config.ini --json
\end{verbatim}
It parses the \texttt{.ini} file, validates the configuration, and reports errors, warnings, and informational messages.

This dry-run step targets a common failure mode in \add{LLM-agent use}.
\add{An LLM-agent} can produce a file that looks plausible but still contains an unsupported scan method, a wrong path, a missing likelihood constraint, a duplicated variable name, an invalid numerical range, or a plot variable that is not defined by any input or output block.
The checker converts such problems into explicit messages that an \add{LLM-agent} can repair and a user can inspect.

It does not decide whether the physics model is correct, but it ensures that the requested scan is at least represented by a syntactically and operationally meaningful \texttt{EasyScan\_HEP} configuration file.

\subsection{Local Web UI}

As shown in Figure~\ref{fig:architecture}, the local Web UI provides an
interactive editor for the same checked \texttt{.ini} configuration used by the
command-line and \add{LLM-agent workflows}.
It allows users to load, edit, preview, check, save, and run
\texttt{EasyScan\_HEP} configurations.
Method-dependent controls are enabled, disabled, or marked inactive according
to the selected scan method; the full non-agent workflow is described in
Section~\ref{sec:run-without-agent}.

\subsection{Machine-readable run interface}

\texttt{EasyScan\_HEP~2} also provides a \add{machine-readable} run interface:
\begin{verbatim}
easyscan --run config.ini --overwrite stop --json
\end{verbatim}
This interface differs from an interactive terminal run in two ways.
First, the overwrite policy is explicit.
If the result folder already exists, the user or \add{LLM-agent} must choose one of
\begin{verbatim}
replace, backup, stop
\end{verbatim}
rather than answering an interactive prompt during execution.
\add{The three choices are inherited from the version-1 interactive prompt for
handling an existing result folder. In \texttt{EasyScan\_HEP~2}, they are
exposed through \texttt{-{}-overwrite} so that non-interactive runs and
LLM-agent workflows can preserve the same behaviour without waiting for terminal
input.}
Second, the run can return a structured report when \texttt{-{}-json} is used.

The structured run report records whether the run succeeded, the return code, the command used to start the run, the launch directory, the configuration path, the log path, the result directory, and the overwrite action.
\add{A shortened example of this JSON report is}
\begin{lstlisting}[style=json]
{
  "command": ["python3", "-m", "easyscan_hep.cli", "config.ini"],
  "config_path": "/path/to/config.ini",
  "cwd": "/path/to/workdir",
  "log_path": "/path/to/config.log",
  "ok": true,
  "overwrite": "stop",
  "result_dir": "/path/to/example_random",
  "return_code": 0,
  "results": {
    "columns": ["x", "y", "f", "Chi2"],
    "row_count": 100,
    "plots": [
      {"name": "Color_x_y_f.png", "size": 18432}
    ],
    "best_row": {
      "x": "1.57",
      "y": "0.02",
      "f": "1.01",
      "Chi2": "0.0025"
    }
  }
}
\end{lstlisting}
\add{The exact numerical values and file names depend on the scan, but the
field structure gives an agent or script a deterministic way to locate logs,
result directories, tables, plots, and representative rows.}
The run output is captured into a log file.
When the run succeeds and the result directory exists, a run manifest is also written to the result directory.
The report gives the \add{LLM-agent} a stable status, log path, and result directory to summarize.

\subsection{Structured result summaries}

Another command-line upgrade is the result-reader interface:
\begin{verbatim}
easyscan --results result_folder --json
\end{verbatim}
This command summarizes an existing result directory without rerunning the scan.
It searches for standard \texttt{EasyScan\_HEP} output tables such as
\begin{verbatim}
ScanResult.txt
All_ScanResult.txt
Previous_ScanResult.txt
EMCEEChain.txt
DynestySamples.txt
\end{verbatim}
and also records generated plot files and other output files.
For tabular output, it reads the column names, counts rows, stores a small preview, and identifies a representative best row when a standard metric is available.
For example, it minimizes \texttt{Chi2} or \texttt{-2lnlike} when such columns are present, and maximizes probability-like columns when appropriate.

This interface is useful for both \add{LLM-agents} and scripts.
After a scan finishes, an \add{LLM-agent} can summarize the number of accepted points, list generated plots, identify the main output table, and point the user to the best row or likelihood-related quantities.
The user does not have to rely on the \add{LLM-agent's} interpretation of terminal output. The result summary is produced by a deterministic reader.

\subsection{Summary of the upgrades}

Table~\ref{tab:agent-upgrades} summarizes the main upgrades in \texttt{EasyScan\_HEP~2}.
Together they allow an \add{LLM-agent} to help prepare and operate a scan while leaving the scientific setup in an explicit, checkable \texttt{EasyScan\_HEP} configuration file.
\add{For users familiar with \texttt{EasyScan\_HEP~1}, the version-1 configuration workflow remains the baseline.  \texttt{EasyScan\_HEP~2} keeps this workflow and adds the interfaces and scan methods summarized in Table~\ref{tab:agent-upgrades}.}

\begin{table}[t]
\centering
\caption{Main \texttt{EasyScan\_HEP~2} upgrades relevant to \add{LLM-agent workflows}.}
\label{tab:agent-upgrades}
\footnotesize
\setlength{\tabcolsep}{4pt}
\renewcommand{\arraystretch}{1.12}
\begin{tabularx}{\textwidth}{@{}
  >{\raggedright\arraybackslash}p{0.20\textwidth}
  >{\raggedright\arraybackslash}p{0.36\textwidth}
  >{\raggedright\arraybackslash}X
@{}}
\toprule
Upgrade & User-facing interface & \add{Purpose in LLM-agent workflows} \\
\midrule
Installable command &
\texttt{easyscan config.ini} &
Stable command outside the source tree \\

Dry-run checker &
\texttt{easyscan -{}-check config.ini -{}-json} &
Validate configurations before running scan points \\

Machine-readable runner &
\texttt{easyscan -{}-run config.ini -{}-json} &
Run scans non-interactively with explicit status, logs, and overwrite policy \\

Structured result reader &
\texttt{easyscan -{}-results results\_folder -{}-json} &
Summarize result tables and generated outputs without rerunning the scan \\

Local Web UI &
\texttt{easyscan -{}-ui config.ini} &
Load, edit, check, save, and run the same \texttt{.ini} configuration \\

Agent skill &
Standalone skill repository &
Guide \add{LLM-agents} to generate, check, revise, and run configurations \\

New scan methods &
\texttt{BESTFIT}, \texttt{EMCEE}, \texttt{DYNESTY} &
Extend the sampler set while reusing the same workflow \\
\bottomrule
\end{tabularx}
\end{table}

These upgrades do not change the role of the \texttt{.ini} file.
Instead, they make the \texttt{.ini} file more useful as the interface between the user, the \add{LLM-agent}, external physics programs, and the numerical sampler.
The \add{LLM-agent} can write and revise the file, but the checker validates it, the UI exposes it, the runner records how it was executed, and the result reader summarizes the files that were produced.
This is the practical meaning of \add{supporting LLM-agent parameter-scan workflows} in \texttt{EasyScan\_HEP~2}.

\subsection{\texorpdfstring{\add{Controlled evaluation of \texttt{EasyScan\_HEP} workflows}}{Controlled evaluation of EasyScan HEP workflows}}
\label{sec:skill-evaluation}

\add{The preceding subsections describe the new interfaces and workflow design
qualitatively. We next evaluate their practical effect on the preparation and
execution of representative parameter-scan tasks.}

\add{We compared three conditions. Condition S used \texttt{EasyScan\_HEP},
matched documentation, and the registered agent skill. Condition D used the
same package and documentation without the skill. In condition M, the model
implemented each scan directly without \texttt{EasyScan\_HEP}. The S--D
comparison isolates the contribution of the skill, while M shows the effect of
the common \texttt{EasyScan\_HEP} workflow.}

\add{All conditions used the \texttt{gpt-5.6-terra} model with medium reasoning
effort in the same Codex environment. The benchmark contained 11 tasks covering sampling methods,
external-program mappings, missing-information handling, a two-program physics
workflow, and serial and parallel execution. The complete definitions of tasks
T1--T11, including their requested workflows and distinguishing requirements,
are provided in ~\ref{app:benchmark-tasks}. Each task was repeated three
times per condition, giving 99 isolated runs. No run used human correction, fallback
assistance, controller-triggered retries, or artifacts from another condition.
All outputs were assessed using the same semantic criteria; condition M was
judged by the requested scientific result rather than by
\texttt{EasyScan\_HEP}-specific file or column names.}

\add{End-to-end (E2E) completion required the final artifacts, requested behavior,
and scientific consistency checks to pass. First-execution success records
whether the first scanner or main-script launch completed the task, while task
and failed executions count all launches and unsuccessful launches,
respectively. The Review LOC (lines of code) metric counts the nonempty,
non-comment lines in the final \texttt{.ini} or main implementation file that
require user inspection. One task required safe stopping before execution, so
E2E includes 33 runs per condition and the process metrics include 30 executable
runs.}

\begin{table}[t]
\centering
\caption{\add{Controlled evaluation of \texttt{EasyScan\_HEP} workflows. First-execution success,
task executions, and failed executions refer to the 30 executable runs per
condition. Review LOC is the median over those runs.}}
\label{tab:skill-evaluation}
\small
\setlength{\tabcolsep}{3.5pt}
\begin{tabularx}{\textwidth}{>{\raggedright\arraybackslash}Xrrrrr}
\toprule
Condition & E2E & First execution & Task & Failed & Review LOC \\
 &  & pass & executions & executions & \\
\midrule
S: EasyScan + skill & 33/33 & 25/30 & 40 & 7  & 24.5 \\
D: EasyScan + docs  & 33/33 & 22/30 & 46 & 8  & 24.5 \\
M: Model only       & 31/33 & 20/30 & 77 & 27 & 123.5 \\
\bottomrule
\end{tabularx}
\end{table}

\add{The results are summarized in Table~\ref{tab:skill-evaluation}.
Conditions S and D both completed 33/33 runs. With the skill,
first-execution success increased from 22/30 to 25/30 and the total number of
task executions decreased from 46 to 40. Failed executions and Review LOC were
similar. The skill therefore did not increase the already saturated E2E
completion rate, but it reduced the number of iterations needed to obtain a
correct workflow.
Condition M completed 31/33 runs and required 77 task executions, including
27 failed executions. Its median Review LOC was about five times that of either
\texttt{EasyScan\_HEP} condition. These results indicate that
\texttt{EasyScan\_HEP} reduces implementation and review effort, while the skill
modestly improves how efficiently the model uses this workflow.
The benchmark quantifies execution outcomes, but does not capture several
broader functions of the skill, including organizing an end-to-end workflow,
guiding the setup of required software, and identifying information missing
from a user prompt before configuration generation. }

\section{\texorpdfstring{Running without an \add{LLM-agent}}{Running without an LLM-agent}}
\label{sec:run-without-agent}

\begin{figure}[t]
\centering
\includegraphics[width=0.95\textwidth]{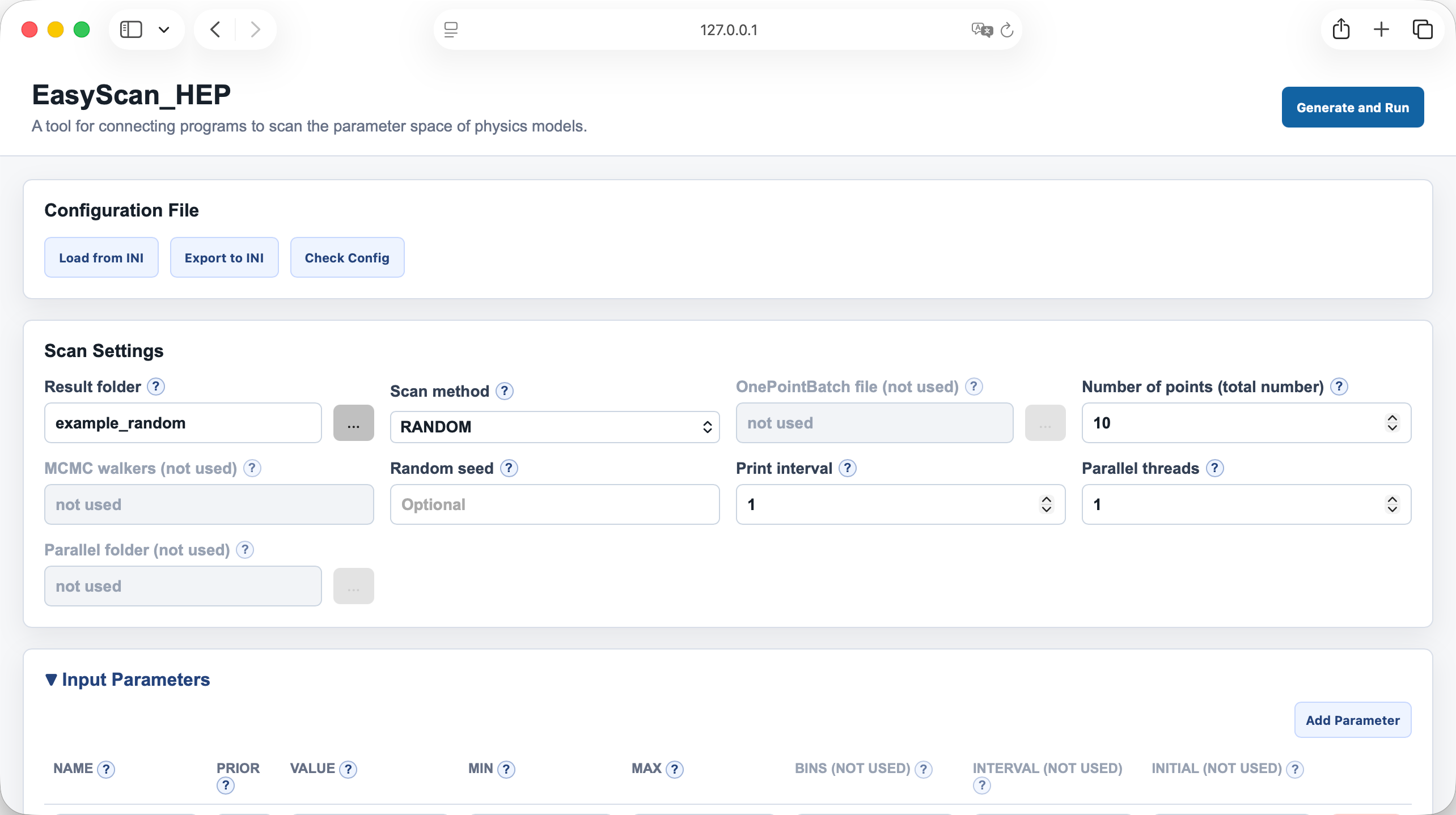}
\caption{Example of the local \texttt{EasyScan\_HEP~2} Web UI for configuration editing.}
\label{fig:local-ui}
\end{figure}

Although \texttt{EasyScan\_HEP~2} supports \add{LLM-agent workflows}, the use of an \add{LLM-agent} is optional.
The conventional \texttt{EasyScan\_HEP} workflow is preserved.
A user may still write or edit a configuration file directly and run it from the command line, for example
\begin{verbatim}
easyscan config.ini
\end{verbatim}
or check it before execution with
\begin{verbatim}
easyscan --check config.ini
\end{verbatim}
\add{These commands operate on the same explicit configuration file used in LLM-agent workflows.}

\texttt{EasyScan\_HEP~2} also provides a local single-user Web UI, which can be launched by
\begin{verbatim}
easyscan --ui
\end{verbatim}
and an existing configuration file can be loaded directly:
\begin{verbatim}
easyscan --ui config.ini
\end{verbatim}
As shown in Figure~\ref{fig:local-ui}, the UI is a visual editor for ordinary \texttt{EasyScan\_HEP} configuration files.
It can build a scan setup, load and save \texttt{.ini} files, call the same dry-run checker used by the command line, launch a scan, display live logs, stop running jobs, open output files, and view generated plots.
The purpose of the UI is to make the file easier to edit and run \add{without delegating the workflow to an LLM-agent}.

When a configuration is imported into the UI, the screen state follows the file.
For example, if the imported file does not contain a \texttt{[constraint]} section, the constraint table is cleared; if it does not contain a \texttt{[plot]} section, the plot table is cleared.
The UI also displays method-dependent configuration fields.
Different scan methods use different options.
For example, \texttt{Bins} is meaningful for \texttt{GRID}; \texttt{Interval} and \texttt{Initial} are meaningful for \texttt{MCMC} and \texttt{EMCEE}; \texttt{MCMC walkers} is specific to \texttt{EMCEE}; and some parallel options are meaningful only for parallel-compatible modes.
This helps the user see which parts of the configuration will affect the backend and which parts will be ignored.

\section{New scan methods}
\label{sec:new-scan-methods}

The configuration-centered design of \texttt{EasyScan\_HEP} makes it straightforward to extend the package with new scan methods.
A scan method in \texttt{EasyScan\_HEP} does not need to define a new way of connecting external physics programs.
It only needs to decide how points are proposed in the parameter space.
The remaining parts of the calculation---writing sampled parameters into input files, running external programs, reading output observables, evaluating constraints, storing results, and producing plots---are already provided by the common \texttt{EasyScan\_HEP} workflow.

\add{In practical terms, a scanner first proposes a trial point in a normalized
parameter space. \texttt{EasyScan\_HEP} maps this point to physical parameters
using the priors specified in the configuration file, runs the external
programs, reads the requested observables, and evaluates the configured
constraints or likelihood.}
The common execution pattern is summarized as
\[
\begin{aligned}
    u \in [0,1]^n
    &\xrightarrow{\;\mathrm{Prior}\;} \theta
    \xrightarrow{\;\mathrm{external\ programs}\;} \mathcal{O}(\theta) \\
    &\xrightarrow{\;\mathrm{constraints}\;}
    \chi^2(\theta)\ {\rm or}\ \ln\mathcal{L}(\theta).
\end{aligned}
\]
Here \(u\) denotes a point in the unit hypercube, \(\theta\) denotes the physical parameters after applying the configured priors, and \(\mathcal{O}(\theta)\) denotes the observables read from external programs.
In \texttt{EasyScan\_HEP~2}, this modular structure was used to add three new scan methods through an \add{LLM-agent-guided} workflow: \texttt{BESTFIT}, \texttt{EMCEE}, and \texttt{DYNESTY}.
The extension route has been encoded in the agent skill.
When a user asks the \add{LLM-agent} to add another scan method, the skill directs it to follow the existing \texttt{BESTFIT}, \texttt{EMCEE}, and \texttt{DYNESTY} pattern, wiring the method through the parser, checker, UI, backend dispatch, example templates, and documentation while keeping the same prior--likelihood interface rather than creating a separate scan driver.

\subsection{\texttt{BESTFIT}: differential-evolution minimization}

The \texttt{BESTFIT} mode is designed for the case in which the immediate goal is to locate a good-fit point rather than to map the full posterior or allowed region.
It minimizes the configured \(\chi^2\) using the differential-evolution optimizer provided by \texttt{SciPy}~\cite{Storn:1997uea,Virtanen:2019joe}.
The optimizer proposes points in the unit hypercube, \texttt{EasyScan\_HEP} transforms them through the configured priors, evaluates the external-program workflow, and returns the resulting \(\chi^2\) to the optimizer.

This mode is useful when a user wants a fast best-fit estimate for a model under a chosen set of observables and constraints.
The \texttt{Number of points} field is interpreted as the maximum number of differential-evolution iterations.
The output follows the usual \texttt{EasyScan\_HEP} result format, so the best-fit point is stored together with the corresponding input parameters, observables, and constraint contributions.

\subsection{\texttt{EMCEE}: ensemble MCMC}

The \texttt{EMCEE} mode adds ensemble Markov-chain Monte Carlo sampling through the optional \texttt{emcee} package~\cite{Foreman-Mackey:2012any}.
It provides a modern multi-walker MCMC backend while preserving the original \texttt{EasyScan\_HEP} configuration style.
Each sampled parameter is specified with its prior, range, proposal interval, and initial value, following the convention already used by the original \texttt{MCMC} mode.

A method-specific field, \texttt{MCMC walkers}, controls the number of walkers.
The checker and backend enforce that this value is compatible with the dimensionality of the scan.
During execution, \texttt{EMCEE} uses the same likelihood function as the other likelihood-driven scan methods.
In addition to the ordinary scan-result files, \texttt{EasyScan\_HEP~2} writes an \texttt{EMCEEChain.txt} file containing the flattened chain information, which can be used for later post-processing.

\subsection{\texttt{DYNESTY}: Python nested sampling}

The \texttt{DYNESTY} mode adds nested sampling through the optional \texttt{dynesty} package~\cite{Speagle:2019ivv}.
It provides a Python-based nested-sampling option that is easier to install in environments where native \texttt{MultiNest} libraries~\cite{Feroz:2008xx} are inconvenient.
As in the other likelihood-driven modes, \texttt{EasyScan\_HEP} supplies the prior transformation and likelihood function, while \texttt{dynesty} controls the exploration of the parameter space.

In this mode, \texttt{Number of points} is interpreted as the number of live points.
The run stores the ordinary scan results and also writes a \texttt{DynestySamples.txt} file containing sampler-specific information such as log-likelihoods, log-weights, evidence-related quantities, and sampled parameters.
This makes the nested-sampling output available without changing the standard \texttt{EasyScan\_HEP} result structure.

\add{Before leaving the scan-method interface, we summarize how the common
\texttt{Number of points} field is interpreted by different backends. For
\texttt{RANDOM}, it sets the number of random trials; for \texttt{MCMC}, it
sets the target number of accepted chain points; for \texttt{BESTFIT}, it sets
the maximum number of differential-evolution iterations; and for
\texttt{MULTINEST} and \texttt{DYNESTY}, it sets the number of live points. For
\texttt{EMCEE}, it is combined with \texttt{MCMC walkers} to determine the
ensemble-chain length. The field is not used by \texttt{GRID}, where the scan
size is instead determined by the per-parameter bin numbers in
\texttt{Input parameters}.}

\section{Physics example}
\label{sec:physics-example}

As a compact phenomenological example, we consider the \(Z_2\)-symmetric
real singlet scalar extension of the Standard Model~(SSM).
The model adds one real scalar field \(S\), odd under \(S\to -S\), coupled to
the Higgs doublet through the Higgs-portal interaction.
At tree level the scalar potential can be written as
\begin{equation}
V_{\rm tree}(H,S)
=
-\mu^2 |H|^2
+\lambda |H|^4
+\lambda_{HS}|H|^2S^2
+\frac{1}{2}\mu_S^2 S^2
+\frac{1}{4}\lambda_S S^4 .
\end{equation}
After electroweak symmetry breaking, the physical singlet mass satisfies
\begin{equation}
m_S^2=\mu_S^2+\lambda_{HS}v_0^2 ,
\end{equation}
with \(v_0\simeq 246~{\rm GeV}\).  In the scan below, \(m_S\) is used as the
physical mass parameter in place of \(\mu_S\), while \(\lambda_{HS}\) is the
portal coupling that controls both the singlet interaction with the Higgs
sector and its finite-temperature contribution to the scalar potential.

The \(Z_2\) symmetry makes \(S\) stable, so that it can act as a
Higgs-portal DM candidate
\cite{Silveira:1985rk,McDonald:1993ex,Burgess:2000yq}.
For thermal freeze-out, the DM relic density is given by
\begin{equation}
\Omega_S h^2=\frac{m_S s_0 Y_S(T_0)}{\rho_c/h^2},
\end{equation}
where  \(Y_S=n_S/s\), \(s_0\) is the current entropy density and \(\rho_c\) is the critical
density.  This example obtains \(\Omega_S h^2\) from
\texttt{micrOMEGAs}~\cite{Alguero:2023zol,Belanger:2026asz}.

At the same time, the Higgs-portal coupling modifies the finite-temperature
effective potential \(V_{\rm eff}(h,s;T)\), which can support a first-order
electroweak phase transition in part of the same parameter space
\cite{Beniwal:2018hyi,Xiao:2022oaq,Balazs:2023kuk,Xiao:2023dbb}.
A cosmological phase transition occurs when the thermal history of the early
Universe drives the scalar background from one phase to another.  Since this
evolution is tracked by the plasma temperature, the critical temperature is
defined by the degeneracy of two phases,
\begin{equation}
V_{\rm eff}(\phi_{\rm false};T_C)
=
V_{\rm eff}(\phi_{\rm true};T_C).
\end{equation}
For the plots below, the transition strength is defined by
\begin{equation}
\frac{v_C}{T_C}
=
\frac{|h_{\rm true}(T_C) - h_{\rm fasle}(T_C) |}{T_C}.
\end{equation}
The example reads \(T_C\), \(h_{\rm true}(T_C)\), and
\(h_{\rm false}(T_C)\) from
\texttt{PhaseTracer}~\cite{Athron:2020sbe,Athron:2024xrh}.

This model is therefore a compact test case for a scan framework.
\texttt{EasyScan\_HEP~2} provides a simple example that performs a two-dimensional grid scan in the plane of \((m_S,\lambda_{HS})\) with the singlet self-coupling fixed to \(\lambda_S=1\).
It first prepares the two external programs from the
source tree by
\begin{lstlisting}[style=ini]
cd EasyScan_HEP
python3 utils/SSM_DM_EWPT/bootstrap.py
\end{lstlisting}
The \texttt{bootstrap.py} script downloads fixed versions of the external
programs, \texttt{micrOMEGAs~7.1} and \texttt{PhaseTracer~2}, applies a
small output-label patch required for reading the \texttt{micrOMEGAs} output,
and compiles both programs.

After the external programs have been prepared, the scan can be performed in three equivalent ways:

\subsubsection*{Direct configuration file}

The first way is the traditional \texttt{EasyScan\_HEP} workflow, in which the
user writes the configuration file directly.
The complete example is provided as
\path|templates/scan_SSM_DM_EWPT.ini|, as shown below:

\begin{lstlisting}[style=ini]
[scan]
Result folder name:  SSM_DM_EWPT_grid
Scan method:       GRID
Input parameters:  mS,        flat,   70,     100,   10
                   lambdaHS,  flat,   0.005,  0.40,  10
Interval of print: 10
Parallel threads:  2
Parallel folder:   utils/SSM_DM_EWPT

[program1]
Program name:    micrOMEGAs_SingletDM
Execute command: ./main data1.par > easyscan_micromegas.out
Command path:    utils/SSM_DM_EWPT/programs/micromegas_7.1/SingletDM/
Input file:      1, utils/SSM_DM_EWPT/programs/micromegas_7.1/SingletDM/data1.par
Input variable:  mS,       1, Label, Mdm1, 2
                 lambdaHS, 1, Label, laSH, 2
Output file:     1, utils/SSM_DM_EWPT/programs/micromegas_7.1/SingletDM/easyscan_micromegas.out
Output variable: Omega_h2,    1, Label, ES_Omega_h2, 2
                 sigmaSIp_pb, 1, Label, ES_sigmaSIp_pb, 2

[program2]
Program name:    PhaseTracer_xSM_TC
Execute command: ./bin/run_xSM_MSbar $(cat easyscan_phasetracer.in)
Command path:    utils/SSM_DM_EWPT/programs/PhaseTracer/
Input file:      1, utils/SSM_DM_EWPT/programs/PhaseTracer/easyscan_phasetracer.in
Input variable:  mS,         1, Position, 1, 1
                 2*lambdaHS, 1, Position, 1, 3
Output file:     1, utils/SSM_DM_EWPT/programs/PhaseTracer/output.txt
Output variable: n_transitions, 1, Position, 1, 4
                 Tc,            1, Position, 1, 5
                 true_h_Tc,     1, Position, 1, 6
                 true_s_Tc,     1, Position, 1, 7
                 false_h_Tc,    1, Position, 1, 8
                 false_s_Tc,    1, Position, 1, 9
[plot]
Color:  mS,       lambdaHS,       Omega_h2,       SSM_Omega_h2
        mS,       lambdaHS,       Tc,       SSM_Tc
\end{lstlisting}

The two external programs use slightly different input conventions.
In the model convention used by the scan and by \texttt{micrOMEGAs}, the portal
coupling is denoted by \texttt{lambdaHS}.
In the \texttt{PhaseTracer} input used here, the corresponding parameter
\texttt{lambda\_hs} differs by a factor of two.
The \texttt{EasyScan\_HEP} configuration therefore passes
\texttt{2*lambdaHS} explicitly to the \texttt{PhaseTracer} input.
The output variables also map directly to the quantities introduced above:
\texttt{Omega\_h2} is the \texttt{micrOMEGAs} value of \(\Omega_S h^2\),
while \texttt{Tc} and \texttt{true\_h\_Tc} give the critical temperature and
Higgs-field value used to form \(v_C/T_C\).

The configuration file can be run using 
\begin{lstlisting}[style=ini]
easyscan templates/scan_SSM_DM_EWPT.ini
\end{lstlisting}
and the syntax follows the \texttt{EasyScan\_HEP~1} configuration manual.

\subsubsection*{Agent skill}

The second way is to prepare the same configuration through the
\texttt{EasyScan\_HEP~2} agent skill \add{in an LLM-agent workflow}.  For this example, the prompt must
specify the parameter ranges, two external programs, convention conversion,
and output variables.  For example, one can ask:
\begin{lstlisting}[style=prompt]
I want a two-dimensional GRID scan for the scalar singlet dark matter model. Scan mS from 70 to 100 and lambdaHS from 0.005 to 0.40, both with flat priors and 10 grid intervals. Use the result folder SSM_DM_EWPT_grid, print every 10 points, and use 2 parallel threads with utils/SSM_DM_EWPT as the parallel folder.

The first external program is micrOMEGAs. Its command path is utils/SSM_DM_EWPT/programs/micromegas_7.1/SingletDM/. Run ./main data1.par and redirect the output to easyscan_micromegas.out. Write mS to the label Mdm1, column 2, and lambdaHS to the label laSH, column 2, in data1.par. Then read Omega_h2 and sigmaSIp_pb from easyscan_micromegas.out using the labels ES_Omega_h2 and ES_sigmaSIp_pb, respectively, column 2.

The second external program is PhaseTracer. Its command path is utils/SSM_DM_EWPT/programs/PhaseTracer/. Run ./bin/run_xSM_MSbar with the contents of easyscan_phasetracer.in as its argument. Write mS to row 1, column 1, and write 2*lambdaHS to row 1, column 3. Then read n_transitions, Tc, true_h_Tc, true_s_Tc, false_h_Tc, and false_s_Tc from output.txt, all from row 1, columns 4 through 9.

Finally, add color plots in the (mS, lambdaHS) plane for Omega_h2 and Tc, named SSM_Omega_h2 and SSM_Tc.
\end{lstlisting}

\subsubsection*{Local Web UI}

The third way is to use the local Web UI.  The user can launch
\begin{lstlisting}[style=ini]
easyscan --ui
\end{lstlisting}
select the \texttt{GRID} method, fill the two scan parameters, add the
\texttt{micrOMEGAs} and \texttt{PhaseTracer} program blocks, and enter the
input--output mappings in the corresponding tables.  An existing template can
also be loaded by
\begin{lstlisting}[style=ini]
easyscan --ui templates/scan_SSM_DM_EWPT.ini
\end{lstlisting}
and then edited through the same fields.

\begin{figure}[t]
\centering
\begin{minipage}[t]{0.48\textwidth}
\centering
\includegraphics[width=\linewidth]{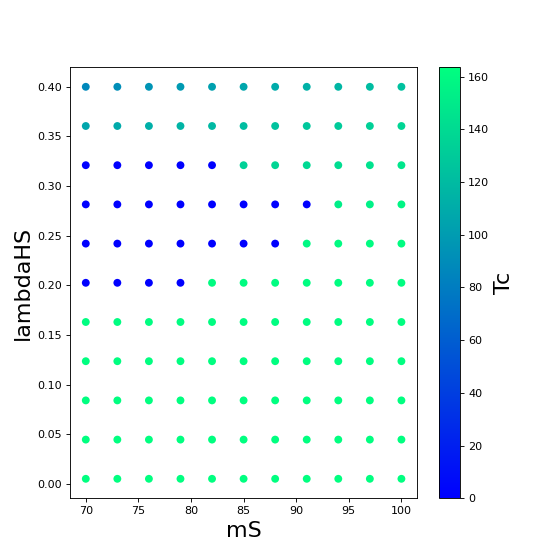}\\[-0.4em]
\small (a) \texttt{EasyScan\_HEP~2} \(T_c\) output.
\end{minipage}
\hfill
\begin{minipage}[t]{0.48\textwidth}
\centering
\includegraphics[width=\linewidth]{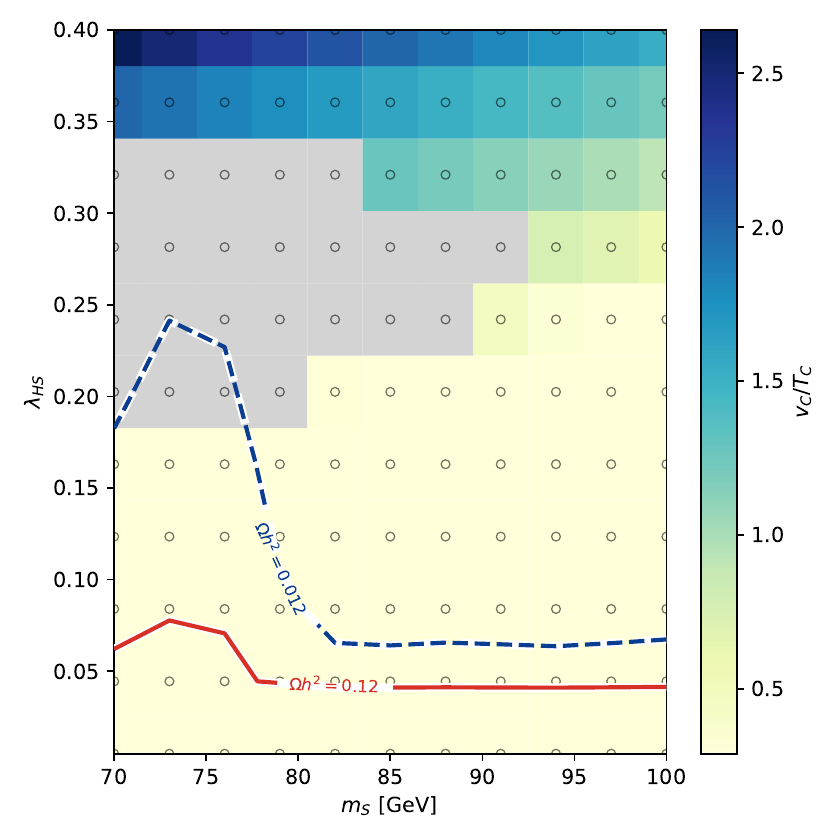}\\[-0.4em]
\small (b) \(v_C/T_C\) with relic-density contours.
\end{minipage}
\caption{Illustration of the scalar-singlet example in the \((m_S,\lambda_{HS})\) plane. Left: the \texttt{EasyScan\_HEP~2}-generated critical-temperature plot. Right: a post-processing plot showing \(v_C/T_C\) together with relic-density contours.}
\label{fig:ssm-dm-ewpt-combined}
\end{figure}

Figure~\ref{fig:ssm-dm-ewpt-combined} illustrates how the same scan result can
be used at two levels.
The left panel is generated directly by \texttt{EasyScan\_HEP~2} from the plot
request in the configuration, and shows the critical temperature returned by
\texttt{PhaseTracer} over the scanned \((m_S,\lambda_{HS})\) plane.
The right panel is a follow-up plot made by asking the \add{LLM-agent} to read the
\texttt{ScanResult.txt} table, compute \(v_C/T_C\) from the
\texttt{Tc} and \texttt{true\_h\_Tc} columns, and overlay relic-density contours.
Because the scan has already been expressed through the agent skill, the
\add{LLM-agent} can carry out this post-processing step from the saved result columns
without the user restating the meaning of the scan or the external-program
mappings.

\section{Conclusion}

\add{As LLM-agent workflows become practical for research use,} \texttt{EasyScan\_HEP~2} adds \add{interfaces for LLM-agent use} to the original \texttt{EasyScan\_HEP} configuration workflow.
The upgraded command-line checker, structured run and result interfaces, local UI, and dedicated agent skill \add{allow an LLM-agent to help prepare, check, run, and summarize configurations} while keeping each scientific calculation tied to an explicit scan description.
The new \texttt{BESTFIT}, \texttt{EMCEE}, and \texttt{DYNESTY} methods broaden the available scan strategies, and their integration shows how future samplers can be added through the same parser, checker, UI, and backend pattern. 
The \add{LLM-agent} does not replace physics validation: generated configurations must
still be inspected by the user, and the checker verifies only syntactic and
operational consistency rather than the correctness of the underlying model or
constraints.

\section*{Acknowledgements}

\add{Generative AI tools were used to assist with the preparation of schematic figures.} This work was supported by the National Natural Science Foundation of China (No. 12335005) and the Natural Science Foundation of Henan Province (No. 262300421233).
The numerical calculations in this work were carried out on the High-Performance Computing Platform at the Center for Theoretical Physics, Henan Normal University.

\clearpage
\appendix
\setcounter{table}{0}
\renewcommand{\theHtable}{A.\arabic{table}}
\section{Benchmark task definitions}
\label{app:benchmark-tasks}

This appendix lists all 11 benchmark tasks used in the controlled
evaluation of Section~\ref{sec:skill-evaluation}. Tasks T1--T5 isolate
sampler-specific requirements using controlled toy
or analytic problems.  Tasks T6--T7 test multi-program workflow construction
and safe handling of incomplete specifications.  Tasks T8--T11 exercise
realistic HEP workflows: T8 integrates dark-matter and phase-transition
calculations, while T9--T11 apply native MCMC, serial random, and two-process
parallel random execution to a common 17-parameter MSSM workflow.
Table~\ref{tab:benchmark-tasks} summarizes the setup, distinguishing
requirement, and primary capability tested by each task.

The MSSM benchmark uses SUSY-HIT for the supersymmetric spectrum and
decays~\cite{Djouadi:2006bz}, FeynHiggs for Higgs-sector
observables~\cite{Hahn:2009zz}, HiggsBounds for collider-exclusion
tests~\cite{Bechtle:2020pkv}, and GM2Calc for the anomalous magnetic moment of
the muon~\cite{Athron:2015rva}.

\begingroup
\footnotesize
\setlength{\tabcolsep}{3pt}
\setlength{\LTcapwidth}{\textwidth}
\renewcommand{\arraystretch}{1.08}
\begin{longtable}{>{\raggedright\arraybackslash}p{0.05\textwidth}>{\raggedright\arraybackslash}p{0.27\textwidth}>{\raggedright\arraybackslash}p{0.34\textwidth}>{\raggedright\arraybackslash}p{0.25\textwidth}}
\caption{Definitions of the 11 tasks used in the controlled evaluation.}
\label{tab:benchmark-tasks} \\
\toprule
Task & Setup and workflow & Distinguishing requirement & Primary capability tested \\
\midrule
\endfirsthead
\caption[]{Definitions of the 11 tasks used in the controlled evaluation (continued).} \\
\toprule
Task & Setup and workflow & Distinguishing requirement & Primary capability tested \\
\midrule
\endhead
\midrule
\multicolumn{4}{r}{\textit{Continued on next page}} \\
\endfoot
\bottomrule
\endlastfoot
\multicolumn{4}{l}{\textit{Controlled sampler semantics}} \\
T1 & A two-dimensional RANDOM scan of \(x\) and \(y\) using a deterministic toy function, with \(f=1.0\pm0.2\) as a Gaussian constraint & Generate 40 trials with the specified seed, apply explicit file-position mappings, and plot both \(f\) and \texttt{Chi2} & Basic end-to-end configuration, execution, and result handling \\
T2 & A two-dimensional GRID scan of \(x\) and \(y\) using a deterministic toy function, with \(f=1.0\pm0.2\) as a Gaussian constraint & Use exactly five equally spaced, endpoint-inclusive values per axis, yielding 25 unique parameter pairs & GRID point-count and endpoint semantics \\
T3 & BESTFIT optimization of a deterministic two-dimensional analytic likelihood & Interpret \texttt{Number of points = 20} as the maximum differential-evolution iteration count and report the best point & Optimizer-specific parameter and objective semantics \\
T4 & EMCEE sampling of a deterministic two-dimensional analytic likelihood & Use eight walkers and 80 target saved samples with the specified initialization, and preserve the chain artifact & Ensemble-MCMC configuration and sampler-specific output \\
T5 & DYNESTY sampling of a deterministic two-dimensional analytic likelihood & Use 50 live points and preserve the sampler-specific samples, log-likelihoods, and weights & Nested-sampling semantics and weighted output \\
\multicolumn{4}{l}{\textit{Workflow construction and safe handling}} \\
T6 & A four-point \((\text{mass},\text{coupling})\) GRID evaluated by two sequential toy calculations & Combine label- and row-position mappings, pass twice the coupling to the second calculation, and merge the two observable outputs & Multi-program orchestration, transformed variables, and mixed mappings \\
T7 & A two-parameter scan requested through an intentionally incomplete external-program interface & Do not inspect or execute the external program and do not invent missing mappings; request focused clarification and stop safely & Defensive handling of underspecified scientific workflows \\
\pagebreak
\multicolumn{4}{l}{\textit{Realistic HEP workflows and execution modes}} \\
T8 & A \(2\times2\) scalar-singlet GRID using micrOMEGAs~7.1 and PhaseTracer~2 & Follow the frozen program mappings and \(2\lambda_{HS}\) convention, and collect both DM and phase-transition observables & Integration of realistic HEP programs in one scientific workflow \\
T9 & The 17-parameter MSSM \(m_W\) workflow and complete SUSY-HIT--FeynHiggs/HiggsBounds--GM2Calc chain with serial native MCMC & Preserve the fixed proposal and acceptance rules and four Gaussian constraints, and obtain 20 accepted points & Faithful reproduction of sampler and likelihood semantics in a complex workflow \\
T10 & The 17-parameter MSSM \(m_W\) workflow and complete SUSY-HIT--FeynHiggs/HiggsBounds--GM2Calc chain with serial RANDOM sampling & Execute exactly 20 independent prior-random trials under a bounded runtime, without posterior or evidence claims & Bounded serial execution of a realistic multi-program workflow \\
T11 & The 17-parameter MSSM \(m_W\) workflow and complete SUSY-HIT--FeynHiggs/HiggsBounds--GM2Calc chain with two-process parallel RANDOM sampling & Split one 20-trial total budget across two real workers with isolated mutable work directories & Parallel execution, worker isolation, and total-budget semantics \\
\end{longtable}
\endgroup
\clearpage

\bibliographystyle{elsarticle-num}
\bibliography{references}

\end{document}